\documentclass[12 pt, article]{article}
\usepackage[left=0.9in, right=0.9in, top=0.75in, bottom=0.75in]{geometry}
\usepackage[utf8]{inputenc}
\usepackage{setspace}
\usepackage{graphicx}
\usepackage{authblk}
\usepackage{sectsty}
\DeclareUnicodeCharacter{2009}{\,}
\usepackage{gensymb}
\usepackage{amsmath}
\usepackage{parskip}
\usepackage{caption}
\usepackage{titlesec}
\usepackage{wrapfig}
\usepackage{chemformula}
\usepackage[utf8]{inputenc}
\graphicspath{{images/}}
\sectionfont{\fontsize{13}{16}\selectfont}

\titlespacing\section{0pt}{12pt plus 2pt minus 2pt}{12pt plus 2pt minus 2pt}
\usepackage[backend=biber, style=chem-acs,articletitle=true,sorting=none]{biblatex}
\title{\Large\textbf {Anomalous interfacial electron transfer kinetics in twisted trilayer graphene caused by layer-specific localization}}

\author[1]{Kaidi Zhang}
\author[1,$\dagger$]{Yun Yu}
\author[2]{Stephen Carr}
\author[3]{Mohammad Babar}
\author[4]{Ziyan Zhu}
\author[1]{Bryan Kim}
\author[1]{Catherine Groschner}
\author[1]{Nikta Khaloo}
\author[5]{Takashi Taniguchi}
\author[6]{Kenji Watanabe}
\author[3]{Venkatasubramanian Viswanathan}
\author[1,7]{D. Kwabena Bediako*}
\affil[1]{\textit{Department of Chemistry, University of California, Berkeley, CA 94720, USA}}
\affil[2]{\textit{Brown Theoretical Physics Center, Brown University, Providence, RI, USA}}
\affil[3]{\textit{Department of Mechanical Engineering, Carnegie Mellon University, Pittsburgh, PA, USA}}
\affil[4]{\textit{SLAC National Accelerator Laboratory, Stanford, CA, USA}}
\affil[5]{\textit{International Center for Materials Nanoarchitectonics, National Institute for Materials Science, Tsukuba, Japan}}
\affil[6]{\textit{Research Center for Functional Materials, National Institute for Materials Science, Tsukuba, Japan}}
\affil[7]{\textit{Chemical Sciences Division, Lawrence Berkeley National Laboratory, Berkeley, CA 94720, USA}}
\affil[$\dagger$]{Current affiliation: Department of Chemistry and Biochemistry, George Mason University, Fairfax, VA, USA}
\affil[*]{Correspondence to: bediako@berkeley.edu}

\date{}
\begin{document}
\maketitle

\doublespacing

\section*{Abstract}

Interfacial electron-transfer (ET) reactions underpin the interconversion of electrical and chemical energy. Pioneering experiments showed that the ET rate depends on the Fermi–Dirac distribution of the electronic density of states (DOS) of the electrode, formalized in the Marcus--Hush--Chidsey (MHC) model. Here, by controlling interlayer twists in well-defined trilayer graphene moirés, we show that ET rates are strikingly dependent on electronic localization in each atomic layer, and not the overall DOS. The large degree of tunability inherent to moiré electrodes leads to local ET kinetics that range over three orders of magnitude across different constructions of only three atomic layers, even exceeding rates at bulk metals. Our results demonstrate that beyond the ensemble DOS, electronic localization is critical in facilitating interfacial ET, with implications for understanding the origin of high interfacial reactivity typically exhibited by defects at electrode–electrolyte interfaces.

\newpage

\section*{Introduction}
Interfacial electron-transfer reactions underpin the interconversion of electrical and chemical energy\supercite{seh2017combining,norskov2009towards,hwang2017perovskites}. Within the microscopic theory of electron transfer, the MHC model\supercite{schmickler2010interfacial,bard2022electrochemical, Chidsey1991,henstridge2012marcus,kurchin2020marcus,schmickler2010interfacial,bard2022electrochemical} highlights the dependence of electrochemical rates on the DOS of an electrode, motivating the discovery of new approaches to manipulating the band structure of electrodes as a means of controlling the performance limits of energy conversion and storage devices. However, it is conventionally assumed that the electrode DOS are invariant with energy/overpotential and delocalized\supercite{kurchin2020marcus}. Atomic defects at electrode surfaces provide a striking, albeit challenging to control, example of the pronounced effect of local structural/electronic modifications on interfacial reactivity. Atomic vacancies\supercite{li2016activating}, kinks, and step edges\supercite{jaramillo2007,güell2015redox,unwin2016nanoscale} are typically associated with massively enhanced interfacial reactivity compared to atomically pristine surfaces. The effect of these defects is typically explained in the context of providing increased DOS at energies that are desirable for charge transfer or formation of a surface-bound catalytic intermediate (such as mid-gap states in a semiconducting material\supercite{jaramillo2007,li2016activating}). However, the dangling bonds at such sites would invariably introduce a strong spatial localization of these large electronic DOS. For this reason, beyond the augmented DOS magnitude, we might consider that localization may play a key role in facilitating interfacial ET to the necessarily localized electronic states on the solution-phase molecule/complex/ion. Yet, a systematic experimental examination of the effects of electronic localization on heterogeneous interfacial charge transfer has been intractable owing to the considerable synthetic challenge of constructing pristine electrode materials that would allow a deterministic modulation of this property separate from the overall DOS. 

Azimuthal misalignment of atomically thin layers produces moiré superlattices and alters the electronic band structure, in a manner that is systematically dependent on the interlayer twist angle\supercite{kim2016van, lau2022reproducibility}. The formation of flat electronic bands, particularly at a series of `magic' moiré angles leads to a diversity of correlated electron physics\supercite{cao2018correlated,hao2021electric,chen2019evidence,balents2020superconductivity}. Notably, these flat bands imply a large DOS that is highly localized in real space\supercite{kerelsky2019maximized}. Small-angle twisted bilayer graphene (TBG) exhibits a recently discovered angle-dependent electrochemical behavior\supercite{yu2022tunable}, where outer-sphere ET kinetics can be tuned nearly tenfold simply by varying moiré twist angle, $\theta_m$, between 0\degree\ and 2\degree.

The stacking order of graphene in multilayers strongly alters the resulting electronic properties of the system\supercite{latychevskaia2019stacking,latil2006charge,xu2015direct,zhu2020twisted,li2020global,zhang2022domino,park2020gate,park2021tunable,hao2021electric,fischer2022unconventional}. Where Bernal (ABA-stacked) trilayer graphene displays dispersive bands, rhombohedral (ABC) graphene possesses a non-dispersive, or `flat', electronic band close to the Fermi level, which is responsible for the emergence of correlated electron phenomena at low temperatures\supercite{zhou2021superconductivity,chen2020tunable}.
\begin{figure}[bthp]
\renewcommand{\figurename}{\textbf{Fig.}}
\renewcommand{\thefigure}{\textbf{\arabic{figure}}}
\centering
  \includegraphics[width=0.6\textwidth]{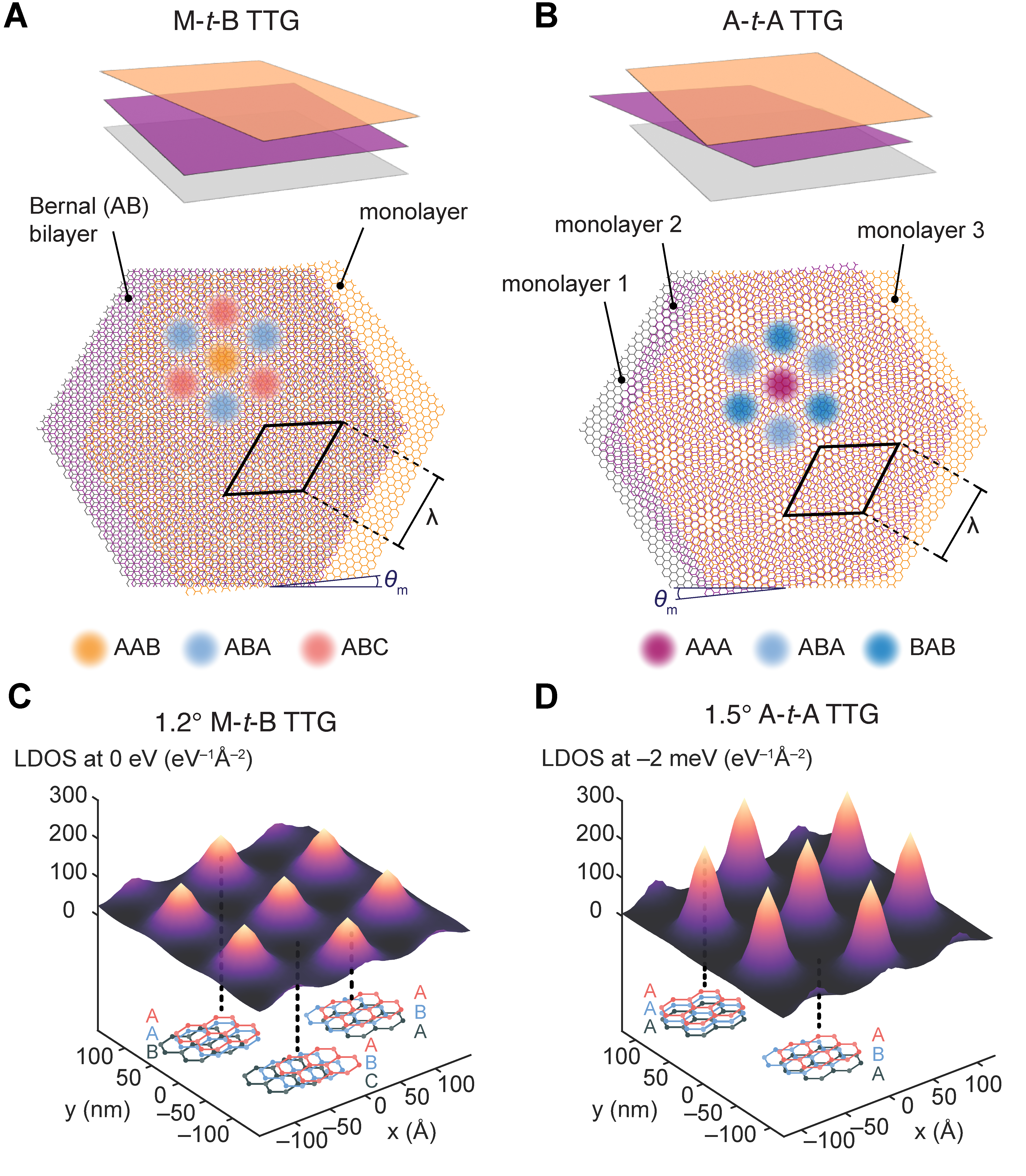}
  \caption{\textbf{Polytypes of twisted trilayer graphene.} \textbf{(A,B)} Illustrations of two twisted trilayer graphene polytypes, with moiré wavelength $\lambda$. The black parallelogram outlines the moiré unit cell in each case. \textbf{(C,D)} Computed local DOS (see Methods) for 1.2\degree\ M-\textit{t}-B (\textbf{C}) and 1.5\degree\ A-\textit{t}-A (\textbf{D}).}
  \label{band structure and DOS}
  \centering
\end{figure} 

More pronounced flat bands are produced in twisted trilayer graphene (TTG) structures. Rotationally misaligned (by a moiré `twist' angle $\theta_m$) monolayer and Bernal stacked bilayer forms a `monolayer-twist-bilayer' (M-\em{t}\em-B) heterostructure (Fig. 1A)\supercite{xu2021tunable,li2022imaging}. Systematically alternating the angle between adjacent graphene layers such that the top layer is perfectly aligned with the bottom layer results in an `A-\em{t}\em-A' heterostructure (Fig. 1B)\supercite{park2021tunable,hao2021electric,fischer2022unconventional} that possesses extremely flat bands at a magic angle around $1.5\degree$. These flattened electronic bands, which manifest as large DOS that are localized on AAB and AAA sites in M-\em{t}\em-B\ and A-\em{t}\em-A TTG, respectively (Figs. 1C,D), now introduce distinctive possibilities for systematically probing the dependence of interfacial ET on electronic structure generally and in particular, the effects of electronic localization. For example, even within the TTG family, larger DOS are found in A-\em{t}\em-A as compared to M-\em{t}\em-B\ near their respective magic angles (Figs. 1C,D), properties that naively might be expected to correlate with interfacial ET rates, based on MHC theory.

\section*{Results and Discussion}
Scanning electrochemical cell microscopy (SECCM)\supercite{unwin2016nanoscale} measurements were carried out on non-twisted (ABA, ABC) and twisted trilayer graphene samples that were fabricated into devices (see Methods)\supercite{yu2022tunable}. As shown in Fig. 2A, naturally occurring ABA and ABC trilayers were mechanically exfoliated from bulk graphite and identified using optical microscopy together with confocal Raman spectroscopy (see Methods) \supercite{lui2011imaging,cong2011raman}. M-\em{t}\em-B\ and A-\em{t}\em-A TTG samples were prepared by the `cut-and-stack' approach (see Methods), resulting in samples possessing uniform $\theta_m$ around the `magic angles' of about 1.34\degree\ for an M-\em{t}\em-B device and 1.53\degree\ for an A-\em{t}\em-A device. Piezoelectric force microscopy (PFM) and scanning tunneling microscopy (STM) were used to evaluate the twist angle distribution and uniformity across the moiré samples (Fig. 2B)\supercite{mcgilly2020visualization}. Using SECCM, cyclic voltammograms (CVs) were measured with 2.0 mM \ch{Ru(NH3)6^{3+}}—an ideal and well-established redox couple for interrogating outer-sphere ET kinetics\supercite{güell2015redox,yu2022tunable}—and 0.10 M KCl as the supporting electrolyte. In Fig. 2C, a representative set of CVs collected from these different trilayer samples is shown.  We find that the ABA domain of the flake shown in Fig. 2A exhibited the most sluggish rates of \ch{Ru(NH3)6^{3+}} electro-reduction, evinced by a half-wave potential ($E_{1/2}$) of –0.32 V, which is cathodically shifted substantially from the equilibrium potential, $E^{0}$, of –0.25 V for \ch{Ru(NH_3)_6^{3+/2+}} (all potentials are reported relative to the Ag/AgCl quasi-counter/reference electrode). However, the $E_{1/2}$ measured from the CV acquired at region II (ABC domain) of the same flake was –0.27 V, pointing to considerably more facile electroreduction kinetics on the rhombohedral trilayer as compared to the Bernal trilayer. For both TTG samples, reversible CVs with $E_{1/2}$ $\approx$ –0.25 V were obtained, indicative of highly facile electrokinetics and heterogeneous electrochemical rate constants that exceed those of both ABA and ABC graphene considerably. These observations motivated the measurement of the variation of interfacial ET rates with $\theta_m$. 
\begin{figure}[hbtp]
\renewcommand{\figurename}{\textbf{Fig.}}
\renewcommand{\thefigure}{\textbf{\arabic{figure}}}
\centering
  \includegraphics[width=1\textwidth]{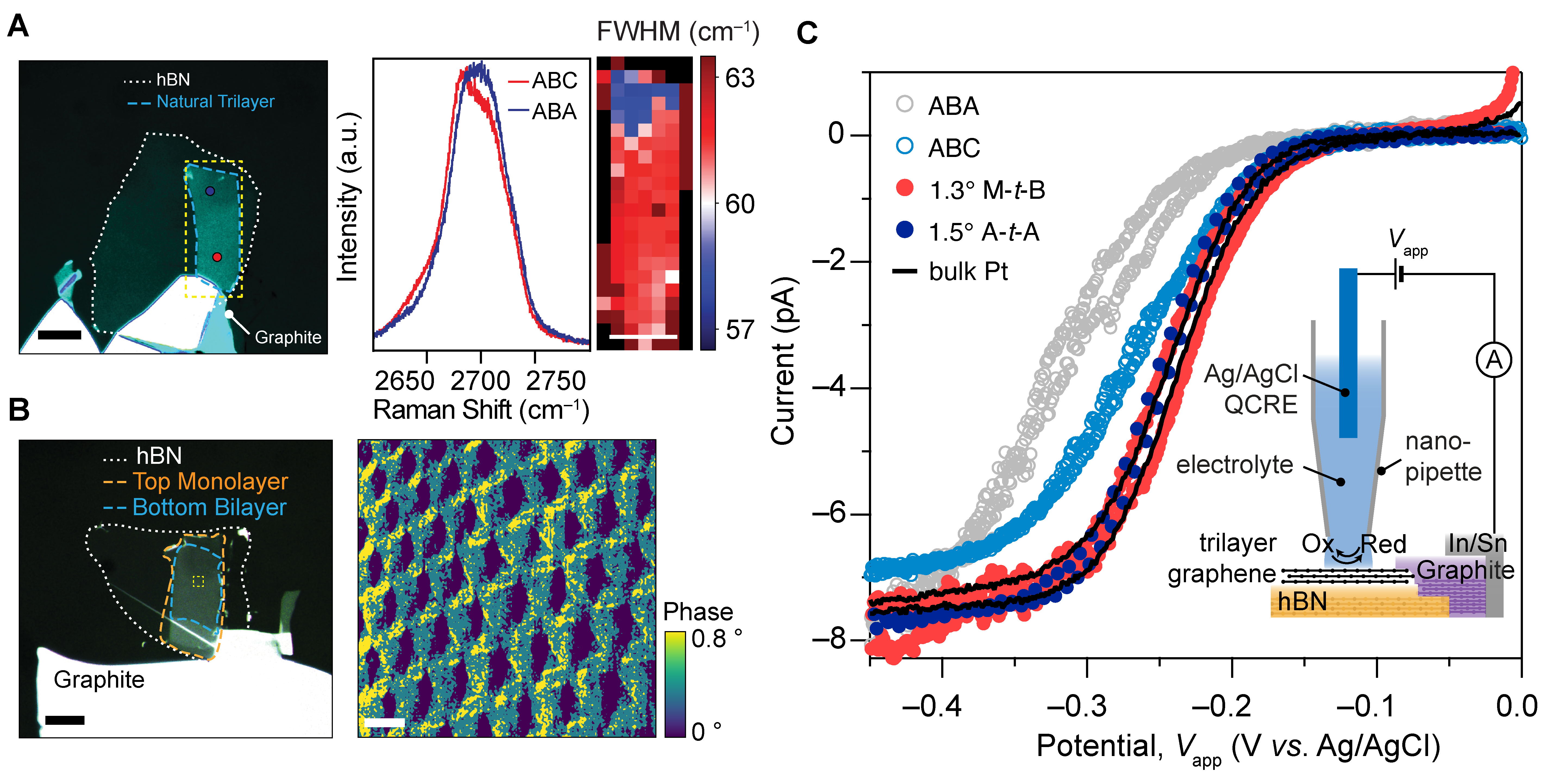}
  \caption{\textbf{Fabrication and electrochemistry of twisted trilayer graphene.} \textbf{(A)} Left: Optical micrograph of a device fabricated from an exfoliated trilayer graphene flake on hBN. Right: Confocal Raman spectra acquired in the sites in \textbf{A} marked with red (ABC domain) and blue (ABA domain) dots, along with the Raman map of the region indicated with a yellow box in \textbf{A}. Scale bars: 10 $\mu$m. \textbf{(B)} Left: Optical micrograph of an M-\textit{t}-B device on hBN (Scale bar: 10 $\mu$m). Right: A lateral PFM phase image over the yellow boxed region in \textbf{B} reveals the moiré superlattice pattern. Scale bar: 50 nm. \textbf{(C)} Representative steady-state voltammograms of 2 mM \ch{Ru(NH3)6^{3+}} in 0.1 M KCl solution obtained at ABA and ABC trilayer graphene, along with 1.3° M-\textit{t}-B and 1.5° A-\textit{t}-A, compared to that obtained at a $\sim 40$ nm thick platinum film. Scan rate, 100 mVs\textsuperscript{-1}. Inset illustrates the SECCM technique.}
  \label{electrochemistry}
  \centering
\end{figure}

To quantitatively assess differences in interfacial kinetics associated with disparate electronic structures, we compared experimental CVs to those simulated with different standard rate constants, $k^0$, calculated with the Butler–Volmer model (see Methods). Here, it is critical to account for the relatively small and potential-dependent quantum capacitance, $C_q$ (see Methods)\supercite{yu2022tunable,güell2015redox} in these low-dimensional electrodes, which for a given applied potential, $V_{app}$, produces a dynamic electron or hole doping of the few-layer graphene by an energy of $eV_q$ (where $e$ is the elementary charge and $V_q$ is the chemical potential relative to the charge neutrality potential). The remainder, $V_{dl}$, persists as a drop across the electric double layer (so that $V_{app} = V_q + V_{dl}$). $C_q(V_q)$ was calculated for all trilayer systems (ABA, ABC, as well as M-\em{t}\em-B and A-\em{t}\em-A at various $\theta_m$) (Fig. 3A) using the respective computed band structures and DOS profiles (see Methods). The corresponding plots of $V_{dl}/V_{app}$ as a function of $V_{app}$ are shown in Fig. 3B. Taken together, these data reveal that flat electronic bands result in a more significant fraction of $V_{app}$ partitioning into $V_{dl}$ near the charge neutrality potential. Notably, as shown in Fig. 3A, changes in $\theta_m$ tune $C_q(V_q)$ and `magic-angle' ($\approx 1.5\degree$) A-\em{t}\em-A displays higher $C_q$ than `magic-angle' ($1.2-1.3\degree$) M-\em{t}\em-B\ consistent with its overall greater DOS (Fig. 1D). 
\begin{figure}[hbtp]
\renewcommand{\figurename}{\textbf{Fig.}}
\renewcommand{\thefigure}{\textbf{\arabic{figure}}}
\centering
  \includegraphics[width=1\textwidth]{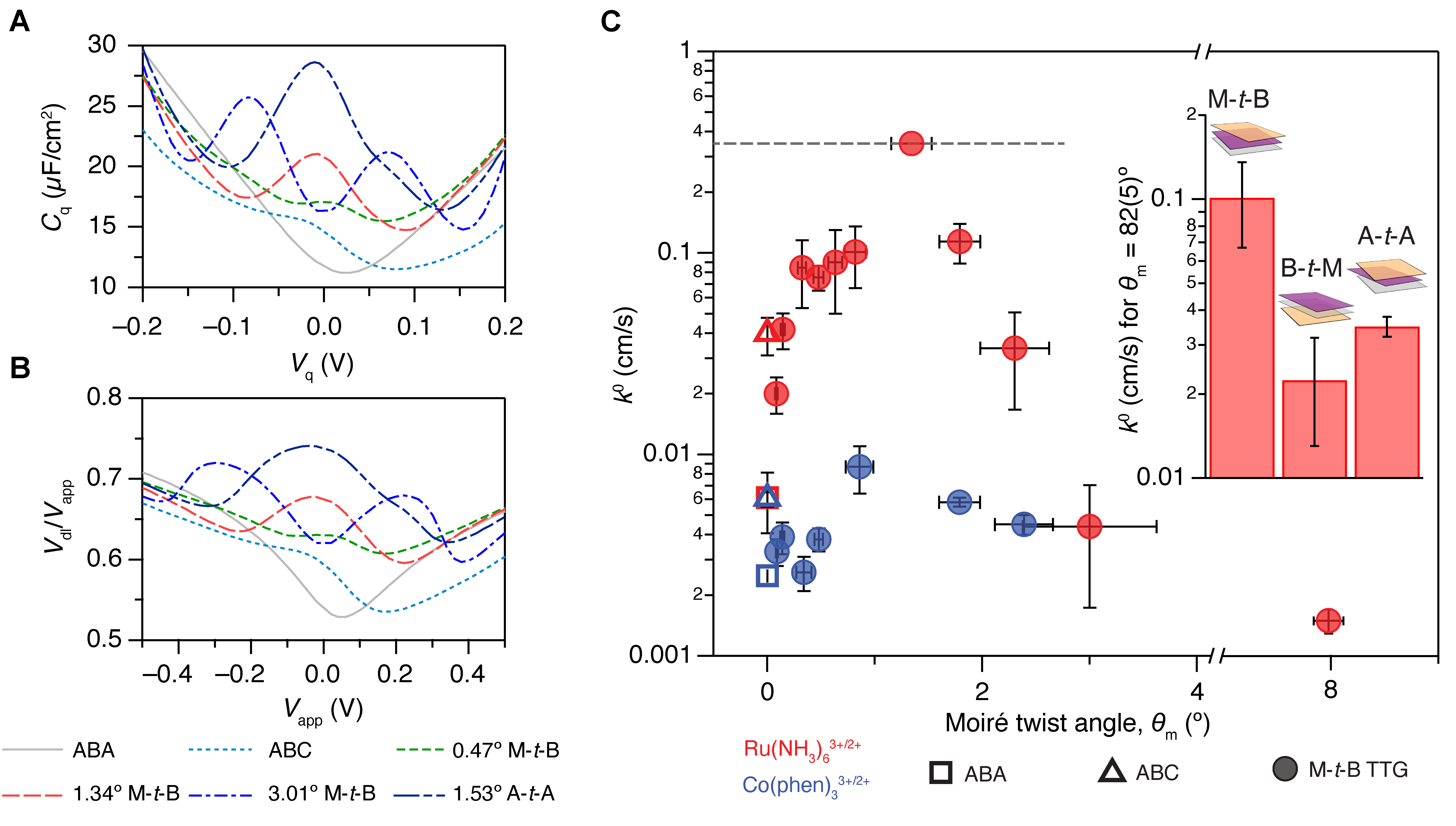}
  \caption{\textbf{Angle dependent quantum capacitance and interfacial ET.} \textbf{(A)} Calculated $C_q$ as a function of the chemical potential ($V_q$) for ABA, ABC, and TTG using the respective computed band structures and DOS profiles (see Methods). \textbf{(B)} Calculated fraction of applied potential on the double layer ($V_{dl}/V_{app}$) as a function of the applied potential ($V_{app}$) for ABA, ABC, and TTG. $V_q$ and $V_{app}$ are relative to the charge neutrality potential. Taken together, these data reveal that flat electronic bands result in a more significant fraction of $V_{app}$ partitioning into $V_{dl}$ near the charge neutrality potential. \textbf{(C)} Dependence of ET rate constant, $k^0$, on trilayer graphene stacking type (ABA, ABC) and $\theta_m$ for M-\em{t}\em-B TTG. Each marker denotes the mean of measurements made on samples within a standard deviation of the mean twist angle. The horizontal and vertical error bars represent the standard deviations of $\theta_m$ and the standard error of $k^0$. The inset shows comparison of $k^0$ values for M-\em{t}\em-B, B-\em{t}\em-M, and A-\em{t}\em-A TTG at $\theta_m = 0.82 \pm 0.05\degree$.}
  \label{electrochemistry}
  \centering
\end{figure}

After determining $V_{dl}$ in this manner, we extracted $k^0$ values by identifying the simulated CV that was in closest agreement with the experiment\supercite{yu2022tunable} (see Methods). The $\theta_m$ dependence of $k^0$ was measured by preparing M-\em{t}\em-B TTG devices with varying $\theta_m$ between 0.08\degree\ and 8.0\degree\ (see Methods) and acquiring CVs of \ch{Ru(NH3)6^{3+}} electroreduction by SECCM for each sample. Fig. 3C shows the strong, non-monotonic variation in $k^0$ over two orders of magnitude from ABA and ABC graphene to $\theta_m =  8\degree$ M-\em{t}\em-B. For samples with $1\degree \le \theta_m \le$ 2\degree\, ET appears reversible within our accessible scan rates and so we cannot extract any kinetic information beyond noting that within this range of $\theta_m$, $k^0 \geq 0.35$ cm/s. The quenched dependence of $\theta_m$ on $k^0$ (blue markers in Fig. 3C) in analogous electrochemical measurements of the  trisphenanthroline cobalt(III/II) redox couple, \ch{Co(phen)_3^{3+/2+}} (see Methods) provides compelling evidence that it is the moiré flat bands that drive the observed angle-dependent electrokinetic modulation in TTG, as in TBG\supercite{yu2022tunable}.

An unexpected observation of the factors controlling interfacial ET is made by comparing the electrochemical responses of TTG polytypes. A-\em{t}\em-A TTG, on the basis of its massive DOS (Fig. 1D) and giant $C_q$—which exceeds that of M-\em{t}\em-B (Fig. 3A)—should be expected to yield the highest ET rates. However, while an effect of $\theta_m$ on $k^0$ is also observed in A-\em{t}\em-A samples, this variant of TTG displays consistently lower $k^0$ than M-\em{t}\em-B at similar $\theta_m$ (Fig. 3C, inset). Further, B-\em{t}\em-M heterostructures, which consist of a Bernal bilayer placed with a twist atop a monolayer (\textit{i.e.}, flipped versions of M-\em{t}\em-B), display markedly lower $k^0$ values than the corresponding M-\em{t}\em-B electrodes, notwithstanding an ostensibly identical overall electronic structure. These striking observations point clearly to effects governing the interfacial ET kinetics beyond simply the ensemble DOS.

\begin{figure}[hbtp]
\renewcommand{\figurename}{\textbf{Fig.}}
\renewcommand{\thefigure}{\textbf{\arabic{figure}}}
\centering
  \includegraphics[width=1\textwidth]{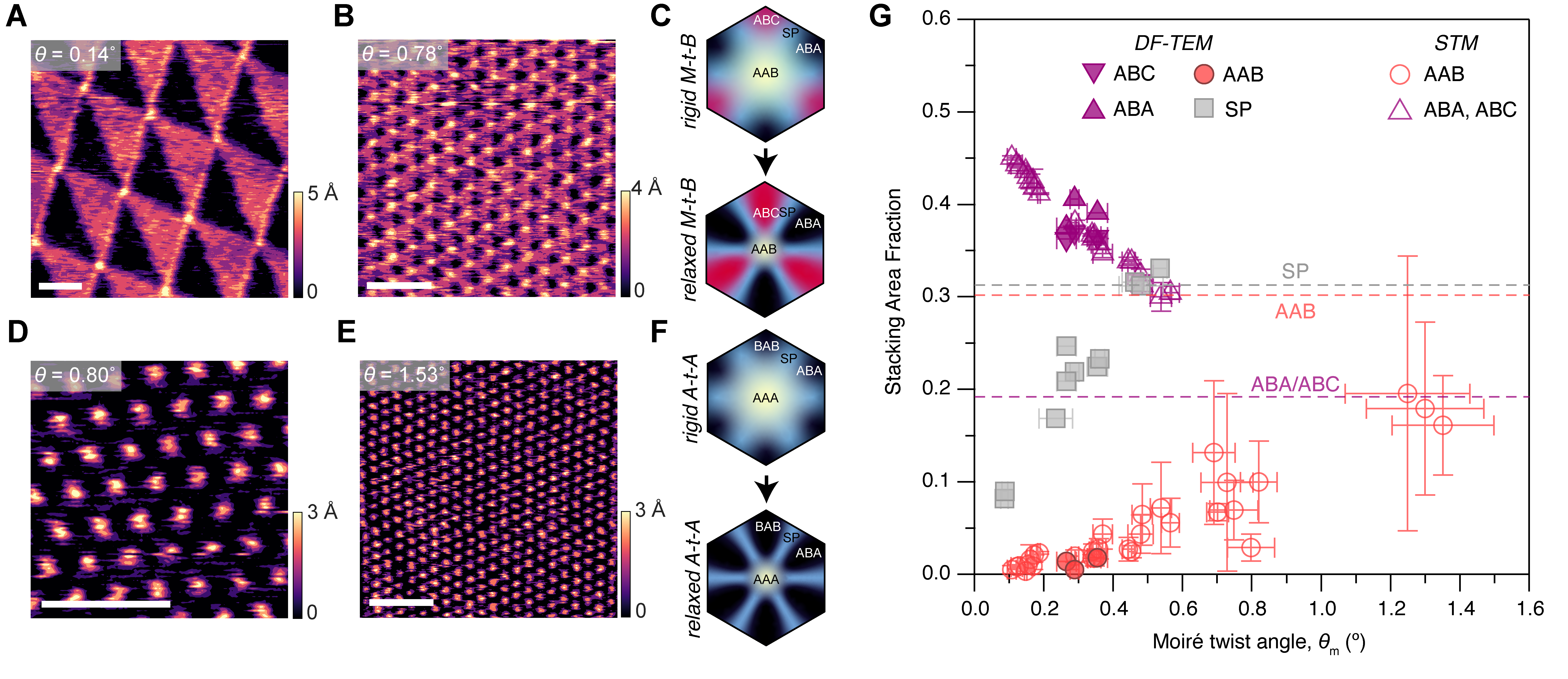}
  \caption{\textbf{Lattice relaxation and stacking area fractions in TTG.} \textbf{(A,B,D,E)} Constant current STM images representative M-\em{t}\em-B (\textbf{A,B}) and A-\em{t}\em-A (\textbf{D,E}) samples. Scale bars: 50 nm \textbf{(C,D)} Qualitative illustrations of different stacking domains in rigid and relaxed M-\em{t}\em-B (\textbf{C}) and A-\em{t}\em-A (\textbf{F}) moiré unit cells. \textbf{(G)} Extracted area fraction of different stacking domains in M-\em{t}\em-B TTG. Error bars represent standard errors.}
  \label{reconstructions}
  \centering
\end{figure}

To fully understand these $\theta_m$ dependencies as well as the disparities between interfacial electron transfer kinetics of M-\em{t}\em-B, B-\em{t}\em-M, and A-\em{t}\em-A, we used STM (room temperature, constant current) to evaluate the role of lattice relaxation in controlling the area fraction of stacking domains in M-\em{t}\em-B and A-\em{t}\em-A TTG. In Fig. 4A, a representative STM map of small-angle ($\theta_m = 0.14\degree$) M-\em{t}\em-B shows a clear contrast between the various stacking domains. Regions with higher local DOS appear brighter than those with lower DOS since a larger tip–sample distance is required to maintain a constant current\supercite{li2022imaging}. ABC domains, therefore, appear brighter than ABA domains owing to the native flatband of the ABC stacking type. These ABA and ABC domains (black and red regions, respectively) form alternating triangular patterns while the AAB region forms small circles of diameter \(\approx\) 11 nm, which appear with the brightest contrast owing to the localization of the moiré flatband and associated large DOS on these AAB sites as shown in Fig. 1C (this is analogous to the localization of moiré flatbands on AA sites in TBG\supercite{kerelsky2019maximized}). For $\theta_m = 0.78\degree$ (Fig. 4B), while the triangular ABA/ABC patterns have shrunk in size compared to those in Fig. 4A, the diameters of AAB regions remained largely unchanged. For A-\em{t}\em-A, AAA domains are visible as bright spots (Figs. 4D,E), consistent with the localization of the large DOS on these regions (Fig. 1D)\supercite{turkel2022orderly},  with degenerate ABA and BAB regions requiring smaller tip–sample distances (dark regions) to sustain a constant STM current because of a lower local DOS.

The measured area distribution of stacking domains in TTG, therefore, differ significantly from those of rigid moiré structures and are instead relaxed as depicted schematically in Fig 4C,F minimizing (maximizing) high (low) energy domains in a manner that is conceptually analogous to that reported for TBG\supercite{yoo2019atomic,kerelsky2019maximized,kazmierczak2021strain}. To support these experiments, we also performed finite element method (FEM) simulations to model relaxation in TTG, finding results that lie in good agreement with our STM and dark-field transmission electron microscopy data. Importantly, these structural measurements and calculations permit a quantitative determination of the area fractions in TTG after reconstruction as a function of $\theta_m$ as plotted in Fig. 4G.

These area fraction distributions after structural relaxation explain the origin of the kinetic modulation observed in Fig. 3C at $\theta_m < 2\degree$ as being driven by $\theta_m$-dependent area fractions of the `topological defect'\supercite{Alden:2013dr,EngelkeTopological} AAB and AAA sites. Our relaxation simulations also show that at $\theta_m \leq 0.3\degree$, relaxation of these moiré superlattices reestablishes nearly commensurate ABA, BAB, and/or ABC domains with local DOS that should not deviate substantially from those of freestanding ABA and ABC trilayers. This observation is in line with previous experimental\supercite{kazmierczak2021strain,yoo2019atomic,li2022imaging,EngelkeTopological} and theoretical studies\supercite{carr2018relaxation,EngelkeTopological} of lattice relaxation in bilayer analogues. Therefore, by considering $k^0$ variations at $\theta_m < 1\degree$ in Fig. 3C (which are also within the range of kinetically resolvable $k^0$), we can extract the local rate constant associated with the AAB and AAA stacking domains through Eq.\ref{mtb} and Eq.\ref{ata} where $\beta_{i}$ and $\kappa_{i}^0$ represent the area fraction and local standard heterogeneous electron transfer rate constant, respectively, for stacking domain $i$. 

\begin{equation}
   k_{MtB}^0 =  \beta_{AAB}\kappa_{AAB}^0 + \beta_{ABC}\kappa_{ABC}^0 + \beta_{ABA}\kappa_{ABA}^0 + \beta_{SP}\kappa_{SP}^0
   \label{mtb}
\end{equation}

\begin{equation}
   k_{AtA}^0 =  \beta_{AAA}\kappa_{AAA}^0 + \beta_{ABA}\kappa_{ABA}^0 + \beta_{BAB}\kappa_{BAB}^0 + \beta_{AAB}\kappa_{AAB}^0
   \label{ata}
\end{equation}

As a result of the lattice relaxation effect discussed above, we can determine $\kappa_{ABA}$ and $\kappa_{ABC}$ from independent measurements of freestanding Bernal and rhombohedral trilayers (Figs. 2C and 3C). In addition, we can assume that $\kappa_{SP}^0 \approx \kappa_{ABA}^0$, which is justified on the basis of the STM images and calculated local DOS. This analysis allows us to extract standard electron transfer rate constants for the AAB (M-\em{t}\em-B), ABB (B-\em{t}\em-M), and AAA (A-\em{t}\em-A) topological defects. 

Combined with previous electrochemical measurements at TBG surfaces\supercite{yu2022tunable} we compare ET kinetics of \ch{Ru(NH3)6^{3+/2+}} among a wide array of stacking configurations from monolayer to trilayer graphene in Fig. 5A. For atomic stacking orders naturally found in bulk graphite, we observed a gradual enhancement as the number of layers increases from monolayer to Bernal trilayer. This can be explained by a modest increase in DOS close to the Fermi level as the number of layers increases.\supercite{güell2015redox} ABC graphene displays a pronounced augmentation in $k^0$ from that of ABA graphene associated with the intrinsic flat band of the rhombohedral system. Most notably, `artificial' high energy stacking (AA, AAA, AAB, ABB) topological defects created by moiré superlattices exhibit extraordinarily high $k^0$ values, with that of AAB exceeding 3 cm/s, which is greater than that measured on bulk platinum electrodes ($0.85–1.2$ cm/s)\supercite{santos1986use}, notwithstanding only consisting of three atomic layers.
\begin{figure}[tp]
\renewcommand{\figurename}{\textbf{Fig.}}
\renewcommand{\thefigure}{\textbf{\arabic{figure}}}
\centering
  \includegraphics[width=1\textwidth]{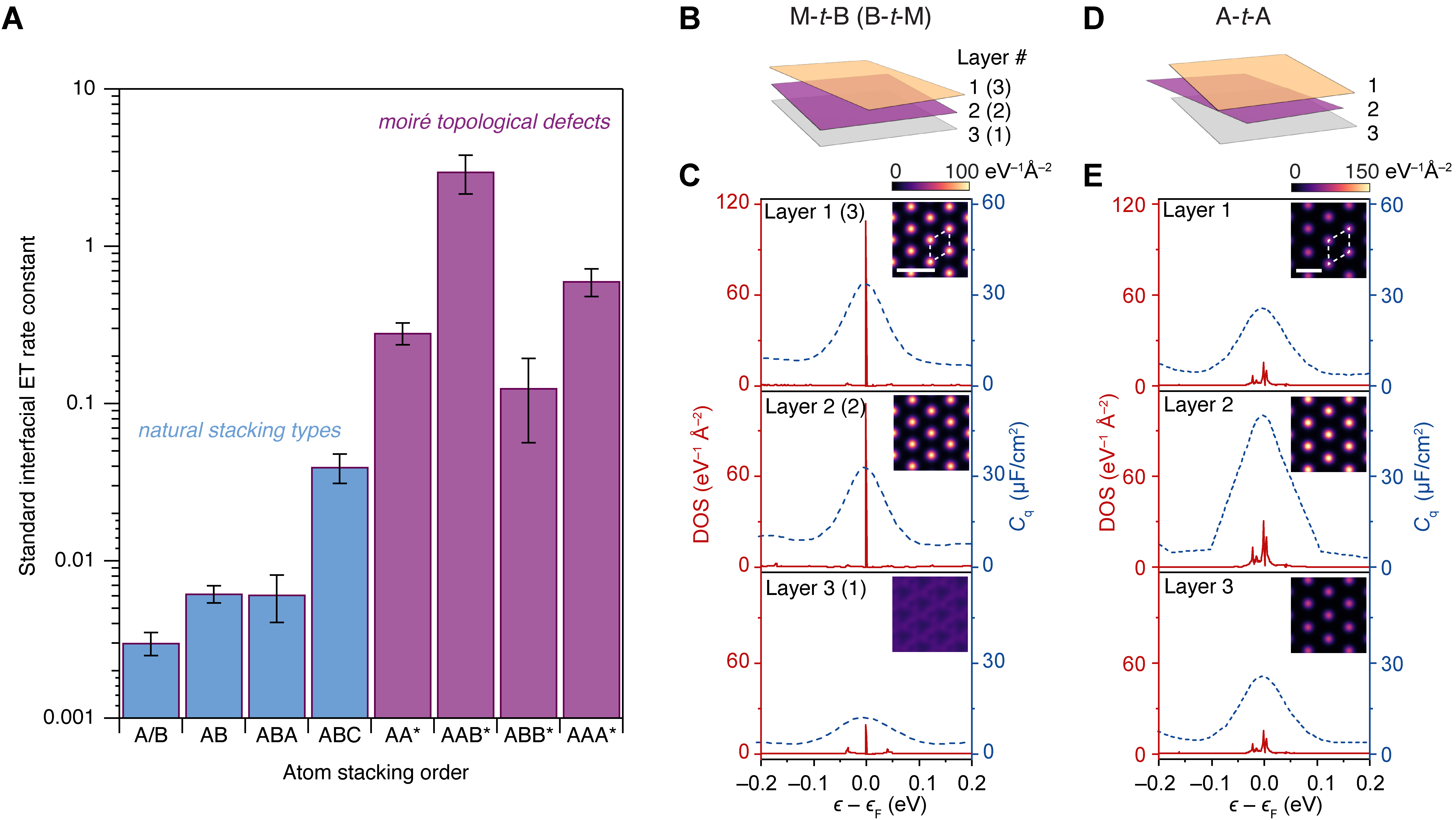}
  \caption{\textbf{ET rates of few-layer graphene and layer-dependent DOS localization} \textbf{(A)} Local standard \ch{Ru(NH3)6^{3+/2+}} ET rate constants at few-layer graphene in different stacking configurations. `Artificial' moiré-derived stacking domains are labeled with an asterisk. Each bar is the mean local rate either measured (for natural stacking) or calculated (for aritifical stacking) for small twist angle samples. The error bars represent the standard errors for the rates. \textbf{(B)} Schematic of M-\em{t}\em-B/B-\em{t}\em-M graphene layers. \textbf{(C)} Layer-dependent DOS profile (see Methods) for AAB stacking domains in M-\em{t}\em-B and B-\em{t}\em-M graphene at $\theta_m = 1.2\degree$. Insets show real space DOS maps of each layer at $\epsilon$ = -3 meV. \textbf{(D)} Schematic of the A-\em{t}\em-A layers. \textbf{(E)} Layer-dependent DOS profile for AAA stacking domains in A-\em{t}\em-A graphene at $\theta_m = 1.2 \degree$. The insets show real space DOS maps of each layer at $\epsilon$ = -1 meV for $\theta_m = 1.2\degree$.}
  \label{Local rates}
  \centering
\end{figure}

Fig. 5A also shows the unexpected result that AAA sites display lower ET rates than AAB notwistanding the higher DOS and $C_q$ of AAA than AAB (Fig. 3A). Strikingly, we also find that ABB sites yield slower ET kinetics than both AAB (despite identical overall DOS) and AA (despite higher overall DOS). Thus, while in-plane electronic localization and structural relaxation effects explain the dependence of $k^0$ on $\theta_m$ in TTG, the relative interfacial ET rates of AAB (M-\em{t}\em-B), ABB (B-\em{t}\em-M), and AAA (A-\em{t}\em-A) (Fig. 3C inset and Fig. 5A) appear not to correlate with DOS. 

To explain these trends, Figs. 5C–E show layer-isolated local DOS($\epsilon$) profiles (Figs. 5C,E) at the topological defects (AAB/ABB, AAA) along with calculated real-space DOS maps (insets in Figs. 5C,E). These calculations show how the DOS enhancements at AAB sites are distinctly localized on the top two layers of M-\em{t}\em-B structures (\textit{i.e.} the `AA' portions of AAB)\supercite{tong2022spectroscopic}. In contrast, the DOS at AAA sites are most strongly localized on the middle layer of A-\em{t}\em-A. This three-dimensional electronic localization (within a thickness of only three atomic layers) arising from different symmetries of these topological defects unveils the fundamental basis for the unexpected trends in ET rate constants at AAB, ABB, and AAA (Figs. 3C and 5A): though the electrodes are only three atomic layers thick, ET rate constants are only correlated with the electronic properties precisely at the electrode--electrolyte interface.

These observations strongly hint at the role of interfacial electronic coupling (between the localized states on the electrode and the electron donor/acceptor in solution), electric double-layer effects, and/or interfacial reorganization energy as even more crucial than the overall DOS alone. Indeed, theoretical calculations based on the MHC model that accounts only for the $\theta_m$-dependent DOS, but with a coupling strength, $\nu$, and reorganization energy, $\lambda$, that are invariant with $\theta_m$, vastly underestimate the dependence of $k^0$ on $\theta_m$. These MHC calculations also likewise predict identical interfacial ET rates for M-\em{t}\em-B and B-\em{t}\em-M, which is clearly at odds with the experiment. Our experimental results, therefore, now motivate future theoretical work to adapt these MHC models to consider how electronic localization, which is deterministically tuned here by varying $\theta_m$ or TTG structure, modifies $\nu$\supercite{pavlov2019role} — and/or $\lambda$\supercite{willard2020solvent}, to bridge the gap between theory and experiment and extend our microscopic understanding of interfacial ET.

\section*{Conclusions}
Controlling stacking geometries and twist angles in few-layer graphene therefore enables the manipulation of standard ET rate constants over three orders of magnitude. In particular, energetically unfavorable topological defects (AAA, AAB stacking domains), which are attainable only through the construction of a moiré superlattice, exhibit extraordinarily high standard rate constants. This electrochemical behavior arises from the moiré-derived flat bands that are localized in these topological defects. In addition to the effects of in-plane structural relaxation and electronic localization, the out-of-plane localization of the electron wavefunction on specific layers of twisted trilayer graphene results in measurable differences in ET rates at topological defects possessing different symmetries. 

These results provide a powerful demonstration of the sensitivity of interfacial ET kinetics to the three-dimensional localization of electronic states at electrochemical surfaces, and raise the question of whether traditional measurements of ET rates at macroscopic electrodes might severely underestimate the true local rate constant, which may be mediated by atomic defects that strongly localize electronic DOS, at these interfaces. In turn, SECCM measurements are shown to be powerful tools for probing layer-dependent electronic localization in atomic heterostructure electrodes. 

Future experimental and theoretical work is needed to shed more light on the microscopic origin of these electron-transfer modulations in the context of reorganization energy, electronic coupling, and even electric double-layer structure. This work also heralds the use of moiré materials as a versatile and systematically tunable experimental platform for theoretical adaptations of the Marcus–Hush–Chidsey framework applied to interfaces with localized electronic states, which are representative of defective surfaces that are ubiquitous to nearly all real electrochemical systems. In an applied context, twistronics is shown to be a powerful pathway for engineering pristine 2D material surfaces to execute charge transfer processes with facile kinetics, holding implications for electrocatalysis\supercite{yu2022Trends,Fu2021MoiréCatalysis} and other energy conversion device schemes that could benefit from ultrathin, flexible, and/or transparent electrodes that retain high electron-transfer kinetics.

\section*{Materials and Methods}
\textit{Chemicals}\\
Natural Kish graphite crystals were purchased from Graphene Supermarket. \ch{Si/SiO2} wafers (0.5 mm thick with 285 nm \ch{SiO2} or 90 nm \ch{SiO2}) were purchased from NOVA Electronic Materials. Polydimethylsiloxane (PDMS) stamps were purchased from MTI Corporation. Sn/In alloy was purchased from Custom Thermoelectric. Poly(bisphenol-A carbonate) (Mw ~45000), dichlorodimethylsilane ($>$99.5\%), hexaammineruthenium(III) chloride (98\%), cobalt(II) chloride hexahydrate (98\%), 1,10-phenanthroline ($>$99\%), calcium chloride ($>$93\%) and potassium chloride ($>$99\%) were purchased from Sigma-Aldrich and used as received. All aqueous electrolyte solutions were prepared with type I water (EMD Millipore, 18.2 M$\Omega$ cm resistivity). The 2 mM solutions of tris(1,10-phenanthroline)cobalt(II) were prepared by dissolving 1:3 molar ratios of solid cobalt(II) chloride and 1,10-phenanthroline in water. In both \ch{Ru(NH3)6^{3+}} and \ch{Co(phen)3^{3+}} solutions, 0.1 M of KCl was added as a supporting electrolyte.

\textit{Sample Fabrication}\\
Graphite and hexagonal boron nitride (hBN) were exfoliated from the bulk crystals with Scotch tape. Exfoliated films were surveyed by an optical microscope (Laxco LMC-5000). Monolayer, bilayer and trilayer graphene were identified with their characteristic optical contrast of ~7\%, 12\% and 18\% respectively in the green channel\supercite{li2013rapid}. Trilayer graphene films were further confirmed by Raman spectroscopy (HORIBA LabRAM Evo) of the 2D peak (around 2600 – 2700 cm\textsuperscript{-1})\supercite{lui2011imaging}. The 2D peak was used to distinguish different stacking domains (ABC/ABA) as ABC trilayer graphene exhibits an enhanced shoulder at around 2640 cm\textsuperscript{-1}. Trilayer graphene and twisted trilayer graphene samples were fabricated by the well-established ‘cut and stack’ dry transfer method\supercite{yu2022tunable}. All transfers were done on a temperature-controlled heating stage (Instec), an optical microscope (Mitutoyo FS70) and a micromanipulator (MP-285, Sutter Instrument). For monolayer twist bilayer or bilayer twist monolayer samples, graphene flakes with both bilayer and monolayer parts were carefully selected. The monolayer section was severed from the bilayer with a scanning tunneling microscopy (STM) tip. For a-twist-a samples, a large piece of graphene ($>$ 50 µm by 20 µm) was cut evenly into three pieces. A thin piece of poly(bisphenol-A carbonate) (PC) film ($\sim$ 3 by 3 mm) attached to a PDMS chunk ($\sim$ 7 by 7 mm) was used to pick up an hBN ($\sim$ 10 – 20 nm) from the \ch{SiO2}/Si substrate at 120 °C. This hBN was carefully aligned with the bottom layer of the graphene stack and lowered to pick up that piece. The stage was rotated (usually to a slightly larger angle than the desired twist) and the second piece of graphene was overlapped by the already picked-up graphene and thus delaminated from the substrate. For a-twist-a samples, a third piece of graphene was picked up after the stage was rotated back to the original orientation. A piece of graphite ($\sim$ 20 nm, $>$ 50 µm by 50 µm) was then picked up such that it was connected to the graphene. The PC film was carefully removed from the PDMS and placed onto a clean \ch{SiO2}/Si. In/Sn was painted onto the graphite via micro-soldering\supercite{girit2007soldering} to a metallic plate which is attached beneath the \ch{SiO2}/Si. 

\textit{Finite Element Simulation and Cyclic Voltammograms Fitting}\\
All finite element simulations of electron transport were performed on COMSOL Multiphysics v5.6 (COMSOL) to capture the effects of quantum capacitance. The fitting of the CVs was achieved by statistical analysis of experimental and simulated CVs.

\textit{Raman Mapping}\\
Confocal Raman spectra were collected by recording from 2550 – 2800 cm\textsuperscript{-1} with a 532 nm laser at 3.2 mW. Raman maps were generated by collecting spectrum across the trilayer films with a step size of 2 µm. The spectrum was fitted with single Lorentzian functions. The full-width half maxima of the fitted functions were used to differentiate ABA and ABC trilayers. 

\textit{PFM Measurements}\\
PFMs were performed on AIST-NT OmegaScope Reflection. Ti/Ir coated silicon probes from Nanosensor with a force constant of ~2.8 Nm\textsuperscript{-1} and resonance frequency of 75 kHz were used. 2 V of AC bias with resonance frequencies at 820 kHz was used and the force was set at 25 nN.

\textit{STM Measurements}\\
STM measurements were conducted using a Park NX10 STM module (Park Systems) at room temperature and atmospheric pressure. Pt-Ir tips were prepared by electrochemical etching of 0.25 mm Pt-Ir wires (Nanosurf) in 1.5 M \ch{CaCl2} solutions\supercite{libioulle1995very}. The scan images were taken with 0.2 V tip-sample bias and 100 pA current set point. Twist angles of various samples were determined using Delaunay triangulation on the Gaussian centers\supercite{yu2022tunable,kerelsky2019maximized}.

\textit{Electron Microscopy Measurements}\\
The transmission electron microscopy images of the nanopipettes were obtained with a JEOl 1200EX transmission electron microscope operated at 100 keV. The top $\sim$ 1 mm portion of the piette was attached to the grid (PELCO Hole Grids) such that the piette tip was positioned in the cetner hole, and the rest of the pipette was broken off. Selected area electron diffraction patterns were collected on a FEI Tecnai T20 S-TWIN transmission electron microscope with a LaB\textsubscript{6} filament operated at 200 kV. Selected area electron diffraction was used to resolve the twist anlges for samples with twist angles larger than $3\degree$. To obtain the diffraction patterns, the fabricated TLG/hBN samples were measured at the National Center for Electron Microscopy facility in the Molecular Foundry at Lawrence Berkeley National Laboratory. Low-magnification DF-TEM images were acquired using a Gatan UltraScan camera on a Thermo Fisher Scientific Titan-class microscope operated at 60 kV.

\textit{Calculation of band structure and DOS}\\
The DOS for trilayer graphene structures was calculated as a function of $\theta_m$ using the ab initio perturbation continuum model developed previously \supercite{carr2019exact}. The low-energy electronic structure is based on a momentum expansion about the valley K point of the super-cell Brillouin zone, allowing a smooth dependence of bands on the twist angle. It has been shown that the perturbation continuum model exactly reproduces the results of the more expensive ab initio tight-binding model, and both are in good agreement with full density functional theory (DFT) calculations \supercite{carr2019exact,carr2020electronic,fang2016electronic,lucignano2019crucial}. The energy range of integration for the DOS was fixed at $\pm$0.5\ eV around the charge neutrality point (CNP). For evaluation of the LDOS, the normalized moiré supercell was divided into a 90 $\times$ 90 grid in real space and sampled over 36 k points in the Brillouin zone. We kept the sublattice symmetry intact and assumed no extra screening of the interlayer coupling constants.

\textit{Quantum Capacitance Calculation}\\
Quantum capacitance ($C_q$) describes the variation of electrical charges with respect to the chemical potential ($V_q$). Theoretical $C_q$ values with respect to $V_q$ was calculated based on the following equation\supercite{yang2015density}:

\begin{equation}
    C_{\mathrm{q}} = e^2\mathop {\int }\limits_{ - \infty }^{ + \infty } D\left( {{\epsilon }} \right)F_{\mathrm{T}}\left( {{{\epsilon }} - eV_{\mathrm{q}}} \right){\mathrm{d}}{{\epsilon }}
   \label{Cq}
\end{equation}

\begin{equation}
F_{\mathrm{T}}\left( {\it{\epsilon }} \right) = \left( {4k_{\mathrm{B}}T} \right)^{ - 1}{\mathrm{sech}}^2\left( {{\it{\epsilon }}/2k_{\mathrm{B}}T} \right)
\end{equation}

where $D(\epsilon)$ is the density of states, which we center at the CNP, $F_T(\epsilon)$ is the thermal broadening function and $kB$ is Boltzmann’s constant. We assumed $T$ = 300 K for our experimental conditions. 
The total electric double-layer capacitance is governed by the compact layer capacitance. Hence, we used a constant $C_{dl}$ = 10 µF cm\textsuperscript{-2} to simplify the calculation \supercite{xia2009measurement}. We solved the self-consistent equations relating $V_{app}$, $V_q$, $V_{dl}$, $C_q$ and $C_{dl}$ using Simpson integration and nonlinear least squares

\begin{equation}
V_{{\mathrm{app}}} = V_{{\mathrm{dl}}} + V_{\mathrm{q}}
\end{equation}

\begin{equation}
\frac{{V_{{\mathrm{dl}}}}}{{V_{\mathrm{q}}}} = \frac{{C_{\mathrm{q}}}}{{C_{{\mathrm{dl}}}}}
\end{equation}

to obtain $C_q$ vs. $V_q$ and $V_{dl}/V_{app}$ vs. $V_{app}$ shown in Fig. 3.

\textit{SECCM Measurements}\\
The SECCM nanopipettes were fabricated from single-channel quartz capillaries (inner and out diameters of 0.7 mm and 1.0 mm from Sutter Instrument) in a laser nano pipette puller (Sutter Instrument Model 2000). The program was set to heat 700, filament 4, velocity 20, delay 127 and pull 140 to generate pipettes of diameters around 200 nm, later confirmed with bright field TEM\supercite{yu2022tunable}. The outer surfaces of the pipettes were silanized by dipping them into dichlorodimethylsilane for 3 - 5 seconds when nitrogen flowed through the inside of the pipettes. They were then filled with either \ch{Ru(NH3)6^{3+}} or \ch{Co(phen)3^{3+}} solutions through a micro syringe. The pipettes were gently tapped and a gentle string of nitrogen was used to eliminate the bubbles. The pipettes were then inserted with a silver AgCl wire as a quasi-counter reference electrode (QCRE). The pipettes were carefully approached (0.2 µm/s) to the locations of interest while a -0.5 V (0.5 V for \ch{Co(phen)3^{3+}}) bias was applied. The meniscus achieved contact when a current of larger than 2 pA (or smaller than -2 pA) was observed. The pipette was allowed to stabilize for 30 s. Cyclic voltammograms (CVs) were then conducted by sweeping the potential at 100 mVs\textsuperscript{-1} between -0.6  to 0 V (0 to 0.8 V for \ch{Co(phen)3^{3+/2+}}) for 5 cycles. Multiple CVs were collected for each sample and for small twist samples ($\theta \le$ 0.15°) with moiré wavelengths of more than 80 nm, only CVs recorded with nanopipettes more than 200 nm in diameter were included to ensure they surveyed multiple stacking domains. To survey electrochemical activities across a large sample, the pipette was retracted by 1 µm after CVs were measured and horizontally moved to a new location for a new approach.

\section*{Acknowledgements}
This material is based upon work supported by the US Department of Energy, Office of Science, Office of Basic Energy Sciences, under award no. DE-SC0021049 (experimental studies by K.Z., Y.Y., B.K., N.K., and D.K.B.) and the Office of Naval Research under award no. N00014-18-S-F009 (computational work by M.B. and V.V.). S.C. acknowledges support from the National Science Foundation under grant no. OIA-1921199. C.G. was supported by a grant from the W.M. Keck Foundation (Award no. 993922). Experimental work at the Molecular Foundry, LBNL was supported by the Office of Science, Office of Basic Energy Sciences, the U.S. Department of Energy under Contract no. DE-AC02-05CH11231. Confocal Raman spectroscopy was supported by a Defense University Research Instrumentation Program grant through the Office of Naval Research under award no. N00014-20-1-2599 (D.K.B.). Other instrumentation used in this work was supported by grants from the Canadian Institute for Advanced Research (CIFAR–Azrieli Global Scholar, Award no. GS21-011), the Gordon and Betty Moore Foundation EPiQS Initiative (Award no. 10637), and the 3M Foundation through the 3M Non-Tenured Faculty Award (no. 67507585). K.W. and T.T. acknowledge support from JSPS KAKENHI (Grant Numbers 19H05790, 20H00354 and 21H05233). We thank Isaac M. Craig for the helpful discussion regarding STM analysis.

\section*{Author Contributions}
K.Z., Y.Y., and D.K.B. conceived the study. K.Z., B.K., C.G., and N.K. performed the experiments.
K.Z. performed the COMSOL simulations. M.B., S.C. and V.V. carried out the theoretical calculations. K.Z. performed the quantum capacitance calculations. K.Z. and B.K. performed STM image analysis. T.T. and K.W. provided the hBN crystals. K.Z., Y.Y., and D.K.B. analysed the data. K.Z. and D.K.B. wrote the manuscript with input from all co-authors.

\printbibliography
\end{document}